\documentclass[sigconf]{acmart}

\usepackage{booktabs} 
\usepackage{graphicx}
\usepackage{graphics}
\usepackage{bbm, dsfont}

\copyrightyear{2018}
\acmYear{2018} 
\setcopyright{iw3c2w3}
\acmConference[WWW 2018]{The 2018 Web Conference}{April 23--27, 2018}{Lyon, France}
\acmBooktitle{WWW 2018: The 2018 Web Conference, April 23--27, 2018, Lyon, France}
\acmPrice{}
\acmDOI{10.1145/3178876.3186130}
\acmISBN{978-1-4503-5639-8/18/04}

\acmArticle{4}

\begin{document}

\title[Me, My Echo Chamber, and I: Introspection on Social Media Polarization]{Me, My Echo Chamber, and I: \\Introspection on Social Media Polarization}

\author{Nabeel Gillani}
\authornote{Authors contributed equally.}
\affiliation{
	\institution{MIT}
}
\email{ngillani@mit.edu}

\author{Ann Yuan}
\authornotemark[1]
\affiliation{
	\institution{MIT}
}
\email{annyuan@mit.edu}

\author{Martin Saveski}
\affiliation{
	\institution{MIT}
}
\email{msaveski@mit.edu}

\author{Soroush Vosoughi}
\affiliation{
	\institution{MIT}
}
\email{soroush@mit.edu}

\author{Deb Roy}
\affiliation{
	\institution{MIT}
}
\email{dkroy@media.mit.edu}

\begin{abstract}
Homophily --- our tendency to surround ourselves with others who share our perspectives and opinions about the world --- is both a part of human nature and an organizing principle underpinning many of our digital social networks.  However, when it comes to politics or culture, homophily can amplify tribal mindsets and produce ``echo chambers'' that degrade the quality, safety, and diversity of discourse online.  While several studies have empirically proven this point, few have explored how making users aware of the extent and nature of their political echo chambers influences their subsequent beliefs and actions.  In this paper, we introduce \textit{Social Mirror}, a social network visualization tool that enables a sample of Twitter users to explore the politically-active parts of their social network.  We use Social Mirror to recruit Twitter users with a prior history of political discourse to a randomized experiment where we evaluate the effects of different treatments on participants' i) beliefs about their network connections, ii) the political diversity of who they choose to follow, and iii) the political alignment of the URLs they choose to share.  While we see no effects on average political alignment of shared URLs, we find that recommending accounts of the opposite political ideology to follow reduces participants' beliefs in the political homogeneity of their network connections but still enhances their connection diversity one week after treatment.  Conversely, participants who enhance their belief in the political homogeneity of their Twitter connections have less diverse network connections 2-3 weeks after treatment.  We explore the implications of these disconnects between beliefs and actions on future efforts to promote healthier exchanges in our digital public spheres.

\end{abstract}


\maketitle

\section{Introduction}
Americans are increasingly sorting themselves according to their ideological stances on political issues \cite{pewPolarization} and allegiances to political parties \cite{primacyPartyism}.  Many prior studies have illustrated how these forces, among others, have contributed to an increase in levels of ``affective polarization'' --- i.e., strong negative emotions members of one particular party feel for those in another \cite{affectNotIdeology}.  This affective polarization is particularly concerning for the quality and nature of civic discourse, as it is often grounded in tribal loyalties and group think that sidestep rational discussion and debate on the issues that extend beyond politics and directly affect quality of life.  

While the historical roots of polarized politics in America run centuries deep \cite{houseDivided}, the rise of digital media and other online discourse platforms have led researchers to investigate how political polarization manifests and evolves in these more modern contexts.  For example, Sunstein highlighted how ``echo chambers'' in online settings --- where individuals with similar political views assemble to discuss particular issues --- can enable mutually-reinforcing opinions to shift individual perspectives towards poles of greater extremity \cite{sunsteinPolarization}.  

Several empirical studies have sought to better-understand how social media platforms can exacerbate polarization.  Yardi and Boyd showed how replies between like-minded Twitter users occur more often than between users who differ in their political views --- and that discussing a highly-politicized issue tends to strengthen group identity \cite{yardiboyd}.  Adamic and Glance uncovered how political blogs tend to link to other blogs that represent similar political ideologies and have few links to those with opposing views \cite{adamicBlogs}.  Conover et al. studied Tweets made during the 2010 mid-term elections to find ideologically-cocooned Twitter retweet networks but ideologically-mixed mention networks --- suggesting homophilous endorsement patterns accompanied by a tendency for politically active users to inject partisan content and ``call out'' their counterparts on the other side \cite{polarizationTwitter}.  Therefore, even when social media platforms create opportunities for cross-cutting dialog, the dialog is often much more about shouting than truly listening.

More recently, Bakshy et al. investigated the nature of political echo chambers on Facebook, where left and right-leaning users not only tend to connect with those who share their perspectives, but also, tend to share different news and other media URLs \cite{bakshyScience}.  While the exact impact of the Internet on mass polarization in America is still up for debate \cite{broadbandPolarization, nberPolarization}, it appears that on platforms like Twitter, users are increasingly following accounts that share their own political views \cite{twitterLongTerm}.

In the face of this ``ideological cocooning'' on social media platforms, it is natural to ask: what, if anything, can --- or even should --- be done about it?  The complexity of human beings' ``moral matrices'' \cite{righteousMind} suggests that simply exposing people to opposing ideological perspectives in hopes of changing their views may actually backfire and further strengthen their \textit{a priori} intuitions and beliefs \cite{debunkingHandbook}.  

Acknowledging this complexity, several interventions have leveraged different strategies to drive socio-political behavior change on social media platforms.  For example, a recent experiment sought to reduce racism on Twitter by programming bot accounts with varying social statuses to publicly scold offending accounts \cite{mungerOnlineRacism}.  Other researchers have designed interventions that harness technology to promote self-reflection and awareness to drive behavior change.  For example, Matias et al. created a web application that showed users how many men and women on Twitter they followed in order to highlight and rectify gender imbalances \cite{followBias}.  To reduce partisan media consumption, Munson et al. built a browser widget that visually reflects back the balance (or lack thereof) in a given users' media consumption habits over some period of time \cite{munsonExtension}.

Despite these examples, to our knowledge, no prior interventions have sought to mitigate political echo chambers by showing users an ideologically-cocooned subset of their digital social networks and asking them to discover their level of social connectedness.  Furthermore, few have measured changes in belief alongside changes in behaviors (e.g. connection diversity and content sharing patterns) that such digital interventions may cause.  Much like prior experiments in the psychological and political sciences have illustrated how providing perspective and exposing ``illusions of understanding'' on political issues can help moderate extreme views \cite{illusionUnderstanding}, we are interested in exploring how guided data visualizations can provide perspective to expose forces that ossify social media echo chambers.  In doing so, we make the following contributions:

\begin{itemize}
	\item A web application that enables a sample of Twitter users to explore their politically-active digital discourse networks,
    \item A proof of concept for a new type of network intervention that explores the impact of data visualizations on participants' beliefs and actions vis-a-vis social media echo chambers, and
    \item A more granular understanding of how social media users perceive digital echo chambers and some of the challenges and opportunities these perceptions may pose for future efforts to mitigate them.
\end{itemize}

\section{Social Mirror} 

Social Mirror is a web application that allows users to explore a sample of their politically-active Twitter connections.  These connections are visually-depicted in the form of a social network.  Our main hypothesis is that by presenting social media users with a bird's-eye view of an ideologically-fragmented social network and asking them to identify their positions within it, we can help cultivate intellectual humility and motivate more diverse content-sharing and information-seeking behaviors.

\subsection{Dataset}

\begin{figure}
\includegraphics[width=1.0\linewidth]{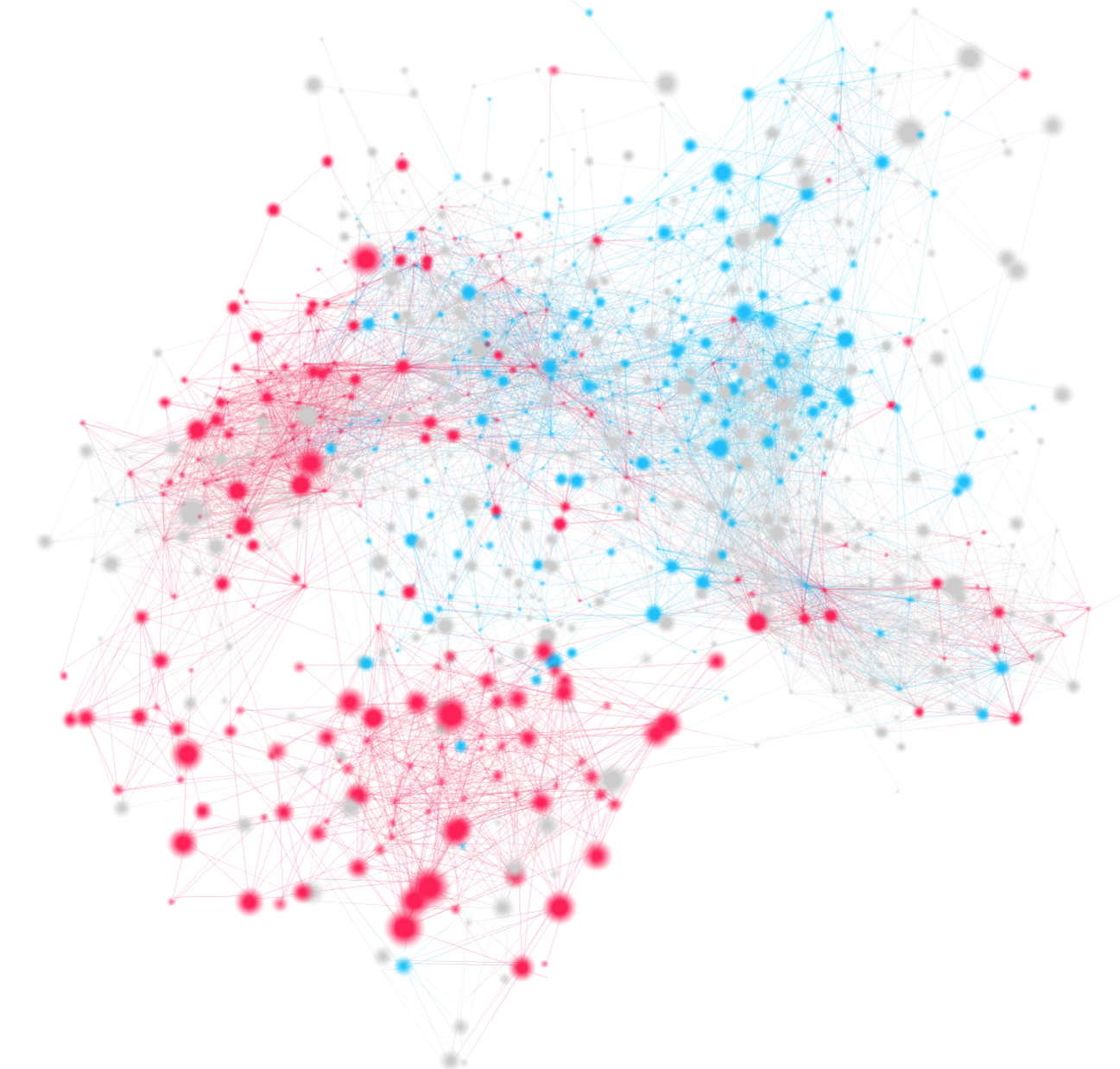}
\caption{\small{Visual depiction of network from the Social Mirror web application.  Nodes represent a sample of nearly 900 Twitter accounts that participated in the conversation about the US Presidential Election between June and mid-September 2016, and edges represent mutual-follower relationships between these accounts. Nodes are sized according to relative PageRank in the depicted network, and colored according to ideology inferred using the political ideology classifier described in \cite{twitterDemos}.  The network is laid out using a standard force directed layout algorithm.  A screen capture video of the application can be found here: \url{https://goo.gl/SRTvxc}.  The live application can be found here: \url{https://socialmirror.media.mit.edu}.  Note that the actual network shown to participants appears on a black background.}}
\label{sm-visual}
\end{figure} 

The network is derived from a sample of 1.1M Twitter users who participated in the conversation about the US Presidential Election on the platform between June and mid-September 2016\footnote{Captured by the Electome project: \url{http://electome.org}.}.  The sample was drawn from a larger project which automatically detected and categorized Election-related Tweets over an 18-month period leading up to the Election \cite{lsmElectome}. 

We built a mutual follower network across this set of users, where a connection between two users means that they followed each other on Twitter at some point within the aforementioned date range.  We then filter down this network to its 4-core --- i.e., the set of users who have at least 4 mutual connections with other users --- yielding approximately 32k nodes and 200k edges.  For visualization, we select approximately 900 of the highest-degree nodes and their corresponding mutual connections among one another.

The social network is visualized using a 3D force directed layout algorithm, which results in accounts with more shared social connections being positioned closer together (Figure \ref{sm-visual}). Interestingly, the visualized network contains two major clusters: one includes a small, tightly-bound set of politically right-leaning accounts, and the other includes a larger set of both right and left-leaning accounts.  Because of the computational expense of dynamically fetching and building the M-degree Twitter social network centered on an arbitrary Twitter user (for some integer M $>$ 1), we limit use of Social Mirror to only those ~32k accounts in the pre-computed 4-core of the full network.

\subsection{User experience}
Upon logging into the application, the user answers a series of questions regarding the nature of her engagement in political discourse on Twitter (details provided in the next section).  Next, the application presents a network visualization of some of the user's immediate Twitter followers and friends.  It then expands to show the entire ~900 node visualization.  The user is encouraged to explore the network by zooming and rotating it.  She can also hover over and highlight groups of accounts to browse which election-related tweets they made during summer 2016.  The application then asks the user to locate herself within the network by clicking on the node that she believes represents her Twitter account. After she makes her guess, the tool reveals her true position and indicates how many degrees (i.e. hops) away it is from her guessed location. The user is also shown a number between 0 and 1, which indicates how politically diverse her connections are.  Depending on which experimental treatment the user is enrolled in, she may also see a list of suggested accounts to follow that would increase her diversity score. Finally, the user is asked to complete a post-survey, which is equivalent to the questionnaire she answered at the beginning of the experience.  The user is also prompted to answer a demographics survey, which asks for her political ideology, age, and gender\footnote{We also asked users to indicate their profession and how they would describe their politics --- i.e. in support of established politics, against established politics, or somewhere in between --- but omit these results from the rest of the study due to data sparsity.}.

\subsection{Technology}
The application is optimized for the Chrome and Firefox web browsers. We used the THREE JavaScript library for WebGL rendering of nodes and edges. We used the JavaScript library Preact --- a simplified version of React --- to manage state on the client. The server is built using the Flask web framework.

\section{Experimental Design}
We are interested in exploring the following overarching question: to what extent does making social media users aware of their online political echo chambers affect their beliefs and future platform engagement patterns?  In particular, we design an intervention to help us measure changes in the following response variables for each participant $p$:

\begin{enumerate}
	\item[\textbf{R1:}] The difference in $p$'s answers to four survey questions administered before and after treatment,
	\item[\textbf{R2:}] The difference in the political diversity of the set of accounts $p$ follows on Twitter before treatment and 1, 2, and 3 weeks after, and
    \item[\textbf{R3:}] The difference in the political alignment of the set of URLs shared by $p$, extracted from the (up to) 200 tweets made before and after treatment.
\end{enumerate}

For R1, the set of pre/post Likert scale survey questions shown to participants include:

\begin{enumerate}
  \item[\textbf{Q1:}] I am well-connected into the conversation about US Politics on Twitter.
  \item[\textbf{Q2:}] Most of my connections on Twitter who discuss politics share my political views.
  \item[\textbf{Q3:}] There are legitimate political views voiced on Twitter that I disagree with.
  \item[\textbf{Q4:}] I would be interested in talking to a Twitter user who does not share my political views.
\end{enumerate}

For R2, we compute political diversity using Shannon's entropy and infer the political ideology of each account $p$ follows using a state-of-the-art classifier presented in \cite{twitterDemos}.  R3 is measured using political alignment scores inferred as a part of \cite{bakshyScience} for different web domains.  The results section provides additional details on each of these outcome variables and how they are computed.  We can think of these response variables as seeking to measure changes in what participants ``believe'' (R1), who they ``listen'' to (R2), and what they ``speak'' about (R3) on Twitter --- all of which help to illuminate the nature of political echo chambers on social media.  

\subsection{Treatments}

We design three treatments in the Social Mirror application and evaluate their effects on the above response variables.  The treatments are designed to gradually increase the amount of information and corresponding set of choices users have in order to mitigate their own echo chambers.  The treatments are as follows:

\begin{description}
  \item[\textit{Viz}:] Participants are shown a mono-color (gray) social network where nodes are not colored according to their ideology.  After participants try and guess their location in the social network, they are shown their true position, along with a diversity score between 0 and 1 representing the political diversity of their connections in the network visualization (where 1 implies a perfectly balanced set of left and right-leaning followers). 
  \item[\textit{Viz+Ideo}:] Same as \textit{Viz}, except nodes are colored according to inferred political ideology (red = right leaning; blue = left leaning; gray = in the middle or unsure).
  \item[\textit{Ideo+Rec}:] Same as \textit{Viz+Ideo}, except users are also recommended up to five Twitter accounts they can follow in order to boost their network political diversity scores.  As users click on accounts to indicate their interest in following them, they are visually shown how following these accounts would increase their connection diversity scores.   
\end{description}
  
\begin{figure*}
\begin{center}
\includegraphics[width=\linewidth]{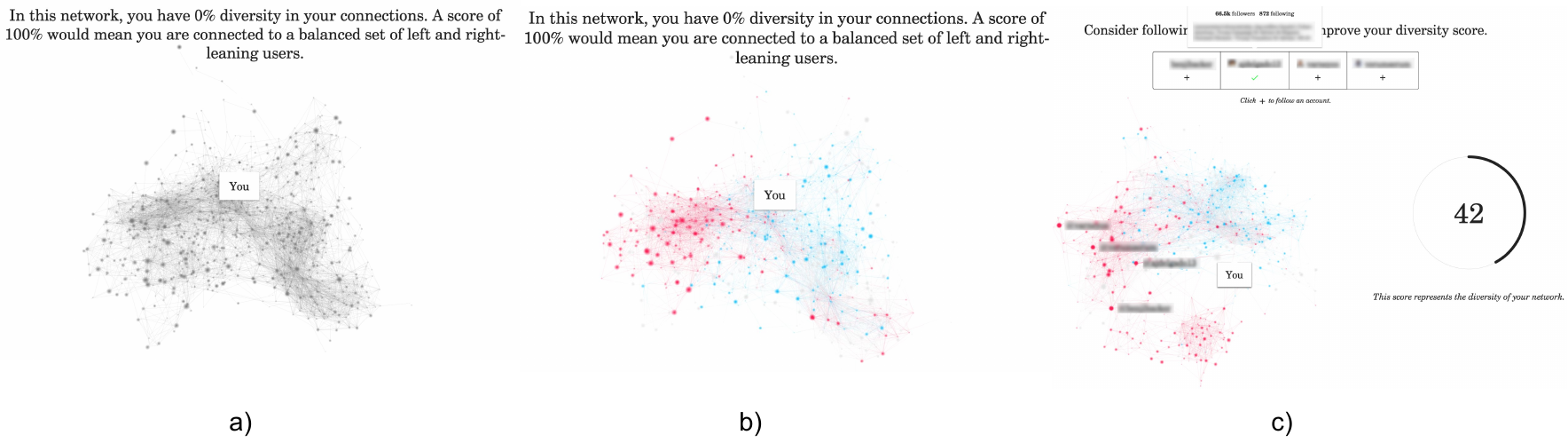}
\caption{\small{Visual depictions of intervention treatment groups: (a) corresponds to treatment \textit{Viz}, where users are asked to find their Twitter account in the network visualization where nodes are not colored by ideology.  Users in \textit{Viz+Ideo} (b) have the same experience as those in \textit{Viz}, but see a social network with nodes colored according to inferred political ideology for each account (red = right-leaning; blue = left-leaning; gray = moderate or unsure).  Users in \textit{Ideo+Rec} (c) have the same experience as those in \textit{Viz+Ideo}, except they are also recommended up to 5 pre-selected Twitter accounts with opposing political views to follow and visually shown how following them would increase their connection diversity scores.  Note that the actual network shown to participants appears on a black background.}}
\label{sm-treatments}
\end{center}
\end{figure*} 

Figure \ref{sm-treatments} shows the three treatments.  In \textit{Ideo+Rec}, accounts are recommended to user $u$ in order of which would have the greatest impact on enhancing $u$'s displayed diversity score.  Therefore, we only recommend accounts that would monotonically increase $u$'s diversity score if followed.  In addition to the treatments, we also sampled 81 users from the original dataset (as a control group for \textbf{R2} and \textbf{R3}) who were never contacted about the study. 

Without knowing the experimental population in advance (i.e. we did not know which of the users we invited would actually participate in the experiment), we were unable to use more sophisticated experimental designs such as completely-randomized or block-randomized assignment. This is a common challenge when running randomized experiments online~\cite{bakshy2014designing}.

\subsection{Recruitment}\label{sec:recruitment}
To recruit study participants, we sampled 1,273 users who had self-enabled their Twitter accounts to accept Direct Messages from any other Twitter account.  We sent them Direct Messages from a research account created for the project, inviting them to participate in the study.  Messages were sent in two rounds --- test and deployment --- to test message quality and select one with the highest likelihood of recruitment (participation results from both rounds are shown together in section 4).  Users who clicked through to the application were randomly assigned to one of the three treatments described above.  We include a copy of the final recruitment message (sent to most users) in the Appendix.  

Users were never required to participate and could choose to completely ignore or disregard the recruitment message.  Upon landing on the application's home page, participants could also read a privacy policy for more information on which kinds of data the application would collect and how it might be used.  This study was approved by MIT's IRB.

\section{Results}
\subsection{Participant statistics}
Approximately 93 users accounted for 108 distinct Social Mirror sessions between June 22, 2017 and August 3, 2017\footnote{I.e., 7.3\% of people who were sent Direct Messages and invited to participate did so in the time horizon highlighted above}.  For the remaining analyses, we consider only data collected during each user's first session.  32, 26, and 35 of the 93 sessions were randomly assigned to experimental treatments \textit{Viz}, \textit{Viz+Ideo}, and \textit{Ideo+Rec}, respectively.  The median length of each session was 5.8 minutes.  Approximately 53 of the 93 unique users completed both the pre and post-experiment survey, and 52 of these 53 answered at least one demographics post-survey question.  Table \ref{sm-treatment-data} shows several user-level properties, including: pre-treatment survey responses, Twitter personas (average URL alignment, connection diversity, verified status, number of followers/followees), and self-reported demographics (gender, age, and political ideology).  Respondents appear to heavily skew male, liberal, and between the ages of 25-44 --- likely influenced by the fact that large portions of Americans are under-represented on Twitter~\cite{twitterNonRepresentative}.   
To test for the robustness of our randomization, we use a logistic regression to check whether there are any significant differences in the distribution of these properties across treatments.  We find no properties have statistically significant coefficients in this model.  We assess the model's overall significance by comparing its log-likelihood to a distribution of log-likelihoods generated by 100k logistic regressions, each of which regresses a random shuffling of participant-treatment assignments against the aforementioned properties of interest~\cite{gerberGreen}.  We find the log-likelihood of our actual model is not statistically significant when compared to this distribution ($p$ = 0.48), suggesting sound randomization and no significant difference in the measured pre-treatment characteristics of the participants.


\begin{table*}
\begin{tabular}{llllllll}
\toprule
\textbf{Property} &  & \textbf{Control} & \textbf{\textit{Viz}} & \textbf{\textit{Viz+Ideo}} & \textbf{\textit{Ideo+Rec}} & \textbf{Totals} \\ 
\midrule
Number of participants &  & 81 & 32 & 26 & 35 & 174 \\ 
Median \# followers (std. dev.) &  & 1562 (9080) & 3023 (27390) & 2471 (9137) & 1724 (29083) &  \\ 
Median \# followees (std. dev.) &  & 711 (350) & 837 (576) & 667 (293) & 717 (424) &  \\ 
$|$Pre URL alignment$|$ (std. dev) &  & 0.184 (0.09) & 0.171 (0.11) & 0.196 (0.07) & 0.180 (0.10) &  \\ 
Pre connection diversity (std. dev) &  & 0.413 (0.17) & 0.428 (0.15) & 0.472 (0.16) & 0.457 (0.136) &  \\ 
Avg Q1 pre-response (std. dev.) &  &  & 4.37 (0.60) & 4.18 (0.73) & 4.29 (0.85) &  \\ 
Avg Q2 pre-response (std. dev.) &  &  & 3.47 (0.9) & 3.24 (1.0) & 3.88 (0.99) &  \\ 
Avg Q3 pre-response (std. dev.) &  &  & 4.26 (0.81) & 4.18 (0.64) & 4.35 (0.61) &  \\ 
Avg Q4 pre-response (std. dev.) &  &  & 3.79 (0.98) & 4.10 (0.56) & 3.94 (1.0) &  \\ 
Verified on Twitter &  & 12 & 8 & 6 & 6 &  32 \\ 
Political ideology &  &  &  &  &  &  \\ 
& Liberal &  & 10 & 7 & 9 &  26 \\ 
& Conservative &  & 2 & 1 & 4 &   7 \\ 
& Moderate &  & 4 & 2 & 0 &   6 \\ 
Gender &  &  &  &  &  &  \\ 
& Female &  & 3 & 4 & 2 &   9 \\ 
& Male &  & 14 & 13 & 13 &  40 \\ 
& Other &  & 1 & 1 & 1 &   3 \\ 
Age &  &  &  &  &  &  \\ 
& 18-24 &  & 4 & 2 & 3 &   9 \\ 
& 25-34 &  & 4 & 4 & 10 &  18 \\ 
& 35-44 &  & 4 & 5 & 3 &  12 \\ 
& 45-54 &  & 2 & 3 & 1 &   6 \\ 
& 55-64 &  & 2 & 3 & 0 &   5 \\ 
& 65+ &  & 2 & 0 & 0 &   2 \\ 
\bottomrule
\end{tabular}
\caption{\small{Twitter persona, pre-survey responses, and demographic information describing Social Mirror experiment participants.  Note that demographic information is only available for 52 out of the 93 participants who were exposed to at least one treatment.  Respondents appear to heavily skew male, liberal, and between the ages of 25-44.  We find no significant imbalances in these values across treatments.}}
\label{sm-treatment-data}
\end{table*}

\subsection{Pre and post-survey responses}

Each question in our pre/post survey can be answered with a response from a 5-point Likert scale ranging from ``Strongly disagree'' (value of 1)  to ``Strongly agree'' (value of 5).  Let $r_{u}^{q_{pre}}$ and $r_{u}^{q_{post}}$ be user $u$'s response to question $q$ in the pre or post-survey, respectively, and $y_{u}^{q}$ = $r_{u}^{q_{post}}$ - $r_{u}^{q_{pre}}$ --- i.e., the change in user $u$'s response to question $q$ after treatment.  We use the following linear regression model to measure the causal effect of each experimental treatment on $y_{u}^{q}$:
\[ y_{u}^{q} = \beta_0 + \beta_{Viz+Ideo} \cdot x_{Viz+Ideo, u} + \beta_{Ideo+Rec} \cdot x_{Ideo+Rec, u} + \epsilon_u, \]

where $x_{Viz+Ideo, u}$ and $x_{Ideo+Rec, u}$ are binary variables that  indicate user $u$'s experimental treatment\footnote{E.g., when the user is in \textit{Viz+Ideo}, $x_{Viz+Ideo, u}$ = 1 and $x_{Ideo+Rec, u}$ = 0.}, $\beta_{Viz+Ideo}$ and $\beta_{Ideo+Rec}$ are their coefficients, respectively, $\beta_0$ is the intercept, and $\epsilon_u$ is the error term.



\begin{table}
\begin{tabular}{lllll} 
\toprule
\ & \textbf{$Q1$} & \textbf{$Q2$} & \textbf{$Q3$} & \textbf{$Q4$}\\
\midrule
$\beta_0$ & -0.32 (0.03)** & 0.11 (0.52) & -0.05 (0.54) & 0 (1) \\
$\beta_{Viz+Ideo}$ & -0.04 (0.86) & 0.42 (0.08)* & -0.01 (0.96) & 0 (1)\\
$\beta_{Ideo+Rec}$ & 0.37 (0.07)* & -0.28 (0.23) & 0.11 (0.37) & -0.18 (0.08)* \\
\bottomrule
\multicolumn{5}{r}{\scriptsize{$^{*}p<0.1$, $^{**}p<0.05$}}
\end{tabular}
\caption{Effects of treatments on changes in between pre and post-responses to each survey question.  Variable coefficients are provided in each cell, with p-values provided in adjacent parenthesis.  N=53.}
\label{sm-survey-deltas}
\end{table}

Table \ref{sm-survey-deltas} shows the average treatment effects of \textit{Viz+Ideo} and \textit{Ideo+Rec} for each of the four questions, relative to 
\textit{Viz}.  Interestingly, the models for Q1, Q3, and Q4 are not statistically significant at the $p$ = 0.1 level - though for Q1, the model is nearly significant ($p$ = 0.11).  Furthermore, for Q1, we find that participants in \textit{Ideo+Rec} tend to increase their belief in how well-connected they are into political discourse on Twitter ($\beta_{Ideo+Rec}$ = 0.37, $p$ = 0.07) compared to those in \textit{Viz}.  When comparing \textit{Ideo+Rec} only to \textit{Viz+Ideo}, the effects are even stronger ($\beta_0$ = -0.35, $\beta_{Ideo+Rec}$ = 0.41, $p$ = 0.03).  We find that both of these results are likely due to the fact that some users in \textit{Ideo+Rec} choose to follow accounts they are recommended towards the end of their treatment. If we remove from the regression those individuals who follow at least one of the recommended accounts (7 out of 35 participants, or 20\%\footnote{To determine if a user has accepted one of our recommendations, we check to see if their list of followed accounts one day after treatment contains at least one of the accounts recommended during treatment.}), the effects of \textit{Ideo+Rec} are reduced with respect to \textit{Viz+Ideo} but still significant at $p$ = 0.1 ($\beta_0$ = -0.35, $\beta_{Ideo+Rec}$ = 0.35, $p$ = 0.08).  Interestingly, this suggests that simply receiving recommendations appears to increase participants' beliefs that they are well-connected.

For Q2 (i.e. \textit{Most of my connections on Twitter who discuss politics share my political views}), we find a relatively strong, significant influence of treatment on participants' beliefs about the level of political homophily among their social ties ($p$ = 0.02 for the overall model).  In particular, participants in \textit{Viz+Ideo} tend to increase their belief that most of their connections on Twitter who discuss politics share their political views ($\beta_{Viz+Ideo}$ = 0.42, $p$ = 0.08).  Looking at pairwise effects between treatments, participants in \textit{Viz+Ideo} increase their belief more than those in \textit{Viz} ($\beta_0$ = 0.11, $\beta_{Viz+Ideo}$ = 0.42, $p$ = 0.09), while participants in \textit{Ideo+Rec} tend to decrease their belief more than those in \textit{Viz+Ideo} ($\beta_0$ = 0.53, $\beta_{Ideo+Rec}$ = -0.71, $p$ = 0.01).  Even after removing those from \textit{Ideo+Rec} who followed recommended accounts, the remaining participants in \textit{Ideo+Rec} are significantly more likely to decrease in their belief that their connections share their political views when compared to those in \textit{Viz+Ideo} ($\beta_0$ = 0.53, $\beta_{Ideo+Rec}$ = -0.61, $p = 0.03$).  Once again, account recommendations seem to decrease participants' beliefs in their level of political homophily on Twitter.

While we see no significant effects of treatments on Q3, we observe several outcomes for Q4.  For one, participants in \textit{Ideo+Rec} decrease in their willingness to have a conversation with a Twitter user who does not share their political views ($\beta_{Ideo+Rec}$ = -0.18, $p$ = 0.08) in the full model with all treatments included --- an effect that strengthens when removing those users in \textit{Ideo+Rec} who follow at least one recommended account ($\beta_{Ideo+Rec}$ = -0.23, $p$ = 0.04).  Pairwise comparisons yield no significant effects. 

Together, these findings suggest an important point: while visually and quantitatively depicting participants' network polarization (\textit{Viz+Ideo}) enhances their belief in the extent to which they live in a political echo chamber, supplementing this with recommendations for people to follow actually produces the opposite effect --- and even leads participants to indicate they are less-inclined to want to have a conversation with someone from the other side of the political aisle.  As recent work has suggested, the choice of account recommendations affects the extent to which political polarization and ``controversy'' decrease on platforms like Twitter \cite{reducingControversy}.  Therefore, under alternative recommendation schemes --- e.g. not seeking to maximize connection diversity --- our findings might differ.

\subsection{Network connection diversity}

\begin{table}[t]
\begin{tabular}{llll} 
\toprule
\ & \textbf{$w = 1$} & \textbf{$w = 2$} & \textbf{$w = 3$} \tabularnewline
\midrule
$\beta_0$ & 0.007 (0.21) & 0.003 (0.64) & 0.01 (0.08)*\\
$\beta_{Viz}$ & -0.008 (0.54) & -0.002 (0.87) & -0.01 (0.52)\\
$\beta_{Viz+Ideo}$ & -0.007 (0.60) & -0.005 (0.75) & -0.01 (0.39)\\
$\beta_{Ideo+Rec}$ & -0.006 (0.66) & -0.003 (0.87) & -0.01 (0.45)\\
\bottomrule
\multicolumn{4}{r}{\scriptsize{$^{*}p<0.1$, $^{**}p<0.05$}}
\end{tabular}
\caption{Effects of treatments on changes in the political diversity of network connections 1-3 weeks after treatment.  Variable coefficients are provided in each cell, with p-values provided in adjacent parenthesis.  N=134 (81 control; 53 across \textit{Viz}, \textit{Viz+Ideo}, and \textit{Ideo+Rec}).}
\label{sm-network-diversity-all}
\end{table}

We also evaluate the effect of treatments on the political diversity of each participant's social network\footnote{Namely, their ``followees'' --- i.e., who they follow on Twitter.} 1, 2, and 3 weeks after use.  In particular, we use Shannon's entropy similar to other studies exploring network connection diversity \cite{networkDiversity,entropyMeasure,measuringBubbles} to quantify the balance of left and right-leaning accounts the user follows.  Under this metric, user $u$'s diversity $w$ weeks after treatment, $d_{u}^{w}$, is~given~by:
\[
	d_{u}^{w} = - \sum_{l \in \mathbb{L}} p_l \log(p_l),
\]
where $\mathbb{L}$ is the set of ideologies (in our case, left or right) and $p_l$ is the fraction of $u$'s followees $w$ weeks after treatment who have ideology $l\in\mathbb{L}$\footnote{Note that our computation of network diversity here is slightly different than the computation used to display to users their connection diversity scores during treatment.  In particular, the displayed scores are computed a) using only their connections in the sampled network of 32k users, not their entire set of followees on Twitter, and b) using terms weighted by the PageRank of the account followed in the sampled network - so that following higher-PageRanking accounts of one ideology versus another leads to a score less than 1 even if the absolute number of connections belonging to either ideology are equivalent.}.  We leverage the political classifier presented in \cite{twitterDemos} to estimate the likelihood of a given Twitter account leaning left or right and use a 60\% threshold to assign final ideology labels\footnote{We use a binary ideological classification scheme for simplicity, though one opportunity for future work includes conducting a similar study with a more flexible, continuous notion of ideology.}.

We explore the effect of \textit{Viz}, \textit{Viz+Ideo}, and \textit{Ideo+Rec} on the change in a given user's connection diversity, compared to a control group of 81 accounts sampled from the same mutual follower network of 32k users but not contacted to participate in the study.  Here, our regression model is:
\begin{align*} 
y_{u}^{w} = \beta_0 & + \beta_{Viz} \cdot x_{Viz, u} + \beta_{Viz+Ideo} \cdot x_{Viz+Ideo, u} \\
  & + \beta_{Ideo+Rec} \cdot x_{Ideo+Rec, u} + \epsilon_u,
\end{align*}

where $x_{Viz, u}$, $x_{Viz+Ideo, u}$, $x_{Ideo+Rec, u}$ are indicator variables for each treatment, $\beta_{Viz, u}$, $\beta_{Viz+Ideo}$, $\beta_{Ideo+Rec}$ are the coefficients representing the corresponding treatment groups, $\beta_0$ is the intercept, $\epsilon_u$ is the error term, and $y_{u}^{w}$ is defined for a given week $w$ as $d_{u}^{w}$ - $d_{u}^{0}$, where $d_{u}^{0}$ is the political diversity of user $u$ immediately preceding her participation in the experiment.  For users in the control group, we let $d_{u}^{0}$ be the diversity of the user's neighborhood as of July 20, 2017 (when we initially recorded their social graphs).  Note that, as a validation filter, we include only treatment users who completed both the pre and post-surveys.  Table \ref{sm-network-diversity-all} shows the results of our model.  We find no significant effects of participating in Social Mirror on a given Twitter user's change in connection diversity 1, 2, or 3 weeks after treatment.  

\begin{table}[t]
\begin{tabular}{llll} 
\toprule
\ & \textbf{$w = 1$} & \textbf{$w = 2$} & \textbf{$w = 3$} \\
\midrule
$\beta_0$ & -0.001 (0.22) & 0.001 (0.54) & 0.002 (0.11)\\
$\beta_{Viz+Ideo}$ & 0.001 (0.38) & -0.003 (0.09)* & -0.004 (0.04)**\\
$\beta_{Ideo+Rec}$ & 0.002 (0.04)** & -0.000 (0.90) & -0.002 (0.23)\\
\bottomrule
\multicolumn{4}{r}{\scriptsize{$^{*}p<0.1$, $^{**}p<0.05$}}
\end{tabular}
\caption{Effects of treatments on changes in the political diversity of network connections 1-3 weeks after treatment.  Variable coefficients are provided in each cell, with p-values provided in adjacent parenthesis.  N=53 (i.e., only those treated users who completed both the pre and post-survey questionnaire).}
\label{sm-network-diversity-survey}
\end{table}


Next, we remove the control group and explore effects across users who participated in one of three treatments.  Again, as a validation filter, we constrain our analyses to only those who also completed the pre and post surveys.  Here, we find several interesting results.  For example, in the week following treatment, participants in \textit{Ideo+Rec} are significantly more likely to have higher network diversity ($\beta_{Ideo+Rec}$ = 0.002, $p$ = 0.04).  Removing those who followed recommended accounts still yields a significant increase, albeit with smaller effects ($\beta_{Ideo+Rec}$ = 0.001, $p$ = 0.08).  In the 2 and 3 weeks that follow treatment, \textit{Viz+Ideo} participants significantly decrease in their connection diversity ($\beta_{Viz+Ideo}$ = -0.003, $p$ = 0.09 and $\beta_{Viz+Ideo}$ = -0.004, $p$ = 0.04, respectively).  Table \ref{sm-network-diversity-survey} summarizes these results.  Looking at pairwise effects between treatments across weeks and removing those in \textit{Ideo+Rec} who followed recommended accounts reveals only one significant effect: a decrease in the connection diversity of \textit{Viz+Ideo} participants compared to those in \textit{Viz} 3 weeks after treatment ($\beta_0$ = 0.002, $\beta_{Viz+Ideo}$ = -0.004, $p$ = 0.07).  

These results suggest that while recommending accounts increases the connection diversity of participants in \textit{Ideo+Rec} in the short-term --- even for those who do not accept these recommendations --- participants in \textit{Viz+Ideo} tend to decrease in their connection diversity 2-3 weeks after treatment compared to those in \textit{Viz}.


\subsection{Political alignment of shared content}

In addition to who social media users choose to follow, the content they choose to share on these platforms also sheds light on their ideological stance \cite{bakshyScience,polarizationTwitter}.  We compute an \textit{average URL political alignment score} for each user $u$, $\bar{A}_{u}$, as follows:
\[ \bar{A}_{u} = \frac{1}{|S|}\sum_{s \in S} A(D(s)), \]

where $S$ is the set of URLs shared by $u$ and $A(s)$ is data from \cite{bakshyScience} which estimates the political alignment of $s$'s web domain, $D(s)$.  A score of -1 for $A(D(s))$ implies URLs that belong to the same web domain as $s$ tend to be shared exclusively by left-leaning social media users; 1 implies these URLs are shared exclusively by right-leaning users; and within this range implies mixed sharing across ideological groups (with 0 indicating balanced sharing).  

Similar to the network diversity analysis, we first explore the effect of any of the treatments on the change in a given user's connection diversity, compared to a control group of 200 Twitter accounts who were sampled from the network but were not contacted to participate in the study.  Here, we use a simple regression model with target variable $y_u$ = $|\bar{A}_{u}^{after}|$ - $|\bar{A}_{u}^{before}|$ --- i.e., the difference in the average political alignment of URLs shared within the $\leq$ 200 most recent tweets made by $u$ before and after treatment (up to 400 tweets total).  Notice that a positive value for $y_u$ indicates an increased tendency for $u$ to share politically-aligned content after treatment compared to before.

We find no significant effects of participation in any treatment group compared to the control.  Next, we look for effects across participants within one of the three treatment groups who also completed both the pre/post-survey.  Once again, we see no significant effects of treatment group on the change in average political alignment of shared URLs ($p$ = 0.7).  Comparing pairwise treatment effects yields a similar null result across pairs.  These null effects are perhaps not surprising given that sharing content on social media is often a sign of public endorsement \cite{sharingEndorsement}, while our study was designed more to explore the effects of reflective engagement with social media echo chambers on what users believe and who they choose to follow.  Still, the absence of effects on average political alignment of shared URLs reveals, in part, that driving different outcomes will require designing different types of interventions.

\section{Discussion}
Our results reveal a disconnect between beliefs and actions: while participants in \textit{Viz+Ideo} are more likely to enhance their belief in the political homogeneity of their Twitter connections, they also exhibit a significant decrease in the actual political diversity of their followees 2-3 weeks after treatment.  Conversely, after treatment, participants in \textit{Ideo+Rec} are more likely to follow Twitter accounts that oppose their political views (driven in part by those who choose to follow the accounts recommended to them as a part of treatment), but also more likely to decrease in their belief that they actually live in a political echo chamber --- and more likely to decrease in their willingness to have a conversation with a Twitter user with opposing political views.  

There are several possible explanations for this disconnect.  For example, immediately after using Social Mirror, \textit{Viz+Ideo} participants may have been convinced by the ideologically-cocooned social network visualization, and their location within it, that their connections are not as politically diverse as they may have previously thought.  But over a slightly longer time horizon, it is possible that users felt that by simply admitting to this imbalance, they were addressing the issue --- and hence, made few additional efforts to actually follow accounts with opposing views or take other actions to address the imbalance.  Participants in \textit{Ideo+Rec} may have received account recommendations that failed to suit their personal preferences, or perhaps they simply did not have enough context or information to know why the recommended account is one worth following --- producing an initial desire to further disengage in discourse with the ``the other'' and a reduced belief in how homogeneous their connections really are.  But despite this belief, many were still nudged into following accounts with opposing political views perhaps simply because the idea of following accounts was presented as an option.

While precise explanations for our results are elusive, it is clear that encouraging discourse across ideological divides is not only a multi-faceted challenge, but also regarded with mixed opinions in the digital public sphere.  We received several thought-provoking written responses to our recruitment messages --- messages that informed recipients about the existence of ideological echo chambers on social media platforms and the importance of engaging with different viewpoints to achieve a ``well-functioning democracy''. For example, some commented that the responsibility of mitigating ideological echo chambers should not simply be in the hands of the individual, citing political institutions and algorithmic curation on social media platforms as sharing responsibility, e.g.:

\begin{quote}
\textit{``Society can not afford to put all the social burden on individuals, while the institutional masses are huddled at Starbucks drinking lattes.''}
\end{quote}

\begin{quote}
\textit{``...it is not necessarily the users. Algorithms social media companies use to put into your recommended feeds rely on similar content to the things you have watched. Most people are very constrained on time and will thus mostly watch what they enjoy, which in political terms means what they agree with. These algorithms pick up on this and then recommend what they agree with.''}
\end{quote}

Others challenged a key assumption underpinning many of our design choices:  ideology as a construct defined along a unidimensional left/right spectrum.  E.g.:

\begin{quote}
\textit{``...many of us have views requiring at least a 2-dimensional representation - we're off the left-right line.''}
\end{quote}

\begin{quote}
\textit{``There are people who have supported Trump for all sorts of reasons. Some people, including myself, wanted Trump to win because we had no idea what he would do. There are other people who wanted Trump to win because they thought he would accelerate the collapse of America. I feel it is very important that you do not use this as an indicator for politics, because virtually everyone I speak with - including myself - dislikes Donald Trump's performance. Most of us, including myself, disliked Donald Trump, but voted for him anyways ... My own politics are hardly represented by the Democrat/Republican dichotomy that is relevant to US politics ... ''}
\end{quote}

Still others challenged the assumption that they were even embedded within a political echo chamber --- or that they should seek out opposing political views:

\begin{quote}
\textit{``I don't at all feel like I am in an echo chamber as your model described - there are no people on Twitter with whom I can agree with on most things, and really it has gotten lonely.''}
\end{quote}

\begin{quote}
\textit{``I systematically avoid anyone who parrots any of the insane talking points of the left. That includes those who want ``fair and balanced'' discussions between insane leftists and sound right. Good luck with your research.I already came to the conclusion that any kind of discussion with the left is a total waste of time.''}
\end{quote}

\begin{quote}
\textit{``All viewpoints are not equally worthy of respect and consideration and I get exposed to enough right wing garbage without you trying to make me seek it out for my own supposed edification.''}
\end{quote}

Despite some users' desires to further disengage, others offered their help or ideas for possible solutions:

\begin{quote}
\textit{``I too am interested in bubbles and actively engaging with different viewpoints. Let me know if I can help!''}
\end{quote}

\begin{quote}
\textit{``There was a Planet money Podcast about a Reddit where there were rules for arguing/stating opposing viewpoints (``On Second Thought'' - June 23, 2017). In it they discussed a civil way to do so. Until that is something that is standard, the online community will continue to funnel itself into echo chambers. Additionally, the amount of faceless cyber bullying on any side of an argument---with no negative recourse, it seems---is yet another reason people of like minds tend to cluster.''}
\end{quote}

The participants in our study and those who responded to our recruitment messages are a biased sample of the American public: Twitter users who actively discuss politics on social media.  Still, our results and their responses, together, reveal some of the challenges and opportunities that await future attempts to mitigate political echo chambers on social media.  

\section{Conclusion}

\noindent
\textbf{Key Findings.}
In this study, we describe an online intervention using \textit{Social Mirror}, a network visualization tool that enables a sample of Twitter users to explore the politically-active parts of their social network.  We invite users to participate in a study in which we vary several software features and ask users to find themselves in an ideologically-cocooned social network in order to assess effects on what participants ``believe'' (changes in their responses to pre/post-survey questions), who they ``listen'' to (who they choose to follow on Twitter after treatment), and what they ``speak'' about (changes in the political alignment of the URLs they share on Twitter after treatment).  Our results reveal a disconnect between belief and action.  Participants who are asked to find their accounts in a sampled social network where nodes are colored by inferred political ideology tend to increase their belief in how ideologically-cocooned they really are, but the political diversity of who they choose to follow on Twitter actually decreases several weeks after treatment.  Conversely, if participants explore an ideologically-cocooned network and are subsequently recommended to follow up to five Twitter accounts with opposing political views, the political diversity of their followees increases one week after treatment --- even though participants believe less that they are in an echo chamber.  

\vspace{3mm}
\noindent
\textbf{Limitations.}
We note several limitations of our study.  For one, our findings are a result of a very short amount of exposure (less than 6 minutes on average) and therefore should not be generalized to contexts where similar treatments may be administered more frequently or over a longer time horizon.  Furthermore, our sample of users is relatively small; recruiting additional users in the future may reduce variance and, perhaps, increase the magnitude of observed effects.  Additionally, flaws in our experimental design prevent us from obtaining a more granular understanding of which parts of the application influenced participants in which ways.  For example, under the current treatments, we are unable to assess the effects of seeing a network visualization (versus just a static score or some other visual indicators of political homophily) on subsequent participants' beliefs and actions.  Additional flaws include potential effects of pre/post-survey question wording and possible network spillover effects on treatment outcomes.  There is also bias embedded in our recruitment efforts: by only contacting those who we could reach through Direct Messages, we may have skewed towards recruiting Twitter personas who are more receptive and/or vocal on social media.  More generally, we are sampling a biased set of Twitter users --- and, of course, the American public as a whole --- by only contacting accounts who participated in 2016 Election-related discourse on the platform during summer 2016, and even more specifically, those who chose to click through and accept the invitation to use the tool.  Even by constraining some of our analyses to only those who completed the pre and post-surveys, we are introducing selection bias that undoubtedly influences the nature and quality of results.  These are important limitations to consider for future interventions.

\vspace{3mm}
\noindent
\textbf{Future Directions.}
There are several opportunities to build on this study.  For one, we may explore additional response variables to further investigate the nature of ideological cocooning on social platforms like Twitter: for example, measuring the impact of future interventions on the quality or civility of subsequent discourse.  The observed disconnect between belief and action also suggests opportunities to design additional measurements that enable us to understand this disconnect and if/how it changes over time.  Given some of the responses to our recruitment messages, we may also design interventions grounded in a more nuanced, multi-dimensional representation of political ideology --- perhaps by focusing on specific topics or issues, instead of creating a global, simplified indicator of participants' political views.  Furthermore, future interventions may peel back the assumption that ``echo chambers are unfavorable'', and instead, seek to understand the individual and environmental factors that produce different opinions on the topic.  Indeed, enabling quality, safe, and diverse discourse in our digital public spheres will require, first and foremost, a degree of humility to listen and learn from all of those who choose to participate.

\section{Acknowledgements}
We would like to thank Peter Krafft for sharing valuable input in the early stages of this project and Prashanth Vijayaraghavan for his instrumental support in accessing the data and classifiers used throughout the study.

\bibliographystyle{ACM-Reference-Format}
\bibliography{sample-bibliography} 


\begin{thebibliography}{28}


\ifx \showCODEN    \undefined \def \showCODEN     #1{\unskip}     \fi
\ifx \showDOI      \undefined \def \showDOI       #1{#1}\fi
\ifx \showISBNx    \undefined \def \showISBNx     #1{\unskip}     \fi
\ifx \showISBNxiii \undefined \def \showISBNxiii  #1{\unskip}     \fi
\ifx \showISSN     \undefined \def \showISSN      #1{\unskip}     \fi
\ifx \showLCCN     \undefined \def \showLCCN      #1{\unskip}     \fi
\ifx \shownote     \undefined \def \shownote      #1{#1}          \fi
\ifx \showarticletitle \undefined \def \showarticletitle #1{#1}   \fi
\ifx \showURL      \undefined \def \showURL       {\relax}        \fi
\providecommand\bibfield[2]{#2}
\providecommand\bibinfo[2]{#2}
\providecommand\natexlab[1]{#1}
\providecommand\showeprint[2][]{arXiv:#2}

\bibitem[\protect\citeauthoryear{{Adamic, L., Glance, N.}}{{Adamic, L., Glance,
  N.}}{2005}]%
        {adamicBlogs}
\bibfield{author}{\bibinfo{person}{{Adamic, L., Glance, N.}}}
  \bibinfo{year}{2005}\natexlab{}.
\newblock \bibinfo{title}{{The political blogosphere and the 2004 U.S.
  election: divided they blog}}.
\newblock \bibinfo{howpublished}{\textit{Proceedings of the 3rd international
  workshop on link discovery}, pages 36-43}.   (\bibinfo{year}{2005}).
\newblock


\bibitem[\protect\citeauthoryear{Bakshy, Eckles, and Bernstein}{Bakshy
  et~al\mbox{.}}{2014}]%
        {bakshy2014designing}
\bibfield{author}{\bibinfo{person}{E. Bakshy}, \bibinfo{person}{D. Eckles},
  {and} \bibinfo{person}{M.S. Bernstein}.} \bibinfo{year}{2014}\natexlab{}.
\newblock \showarticletitle{Designing and deploying online field experiments}.
  In \bibinfo{booktitle}{{\em Proceedings of the 23rd International Conference
  on World Wide Web}}. ACM, \bibinfo{pages}{283--292}.
\newblock


\bibitem[\protect\citeauthoryear{{Bakshy, E., Messing, S., Adamic,
  L.}}{{Bakshy, E., Messing, S., Adamic, L.}}{2015}]%
        {bakshyScience}
\bibfield{author}{\bibinfo{person}{{Bakshy, E., Messing, S., Adamic, L.}}}
  \bibinfo{year}{2015}\natexlab{}.
\newblock \bibinfo{title}{{Exposure to ideologically diverse news and opinion
  on Facebook}}.
\newblock \bibinfo{howpublished}{\textit{Science}, 348 (6239), 1130-1132}.
  (\bibinfo{year}{2015}).
\newblock


\bibitem[\protect\citeauthoryear{{Bateman, D.A., Clinton, J.D., Lapinski,
  J.S.}}{{Bateman, D.A., Clinton, J.D., Lapinski, J.S.}}{2012}]%
        {houseDivided}
\bibfield{author}{\bibinfo{person}{{Bateman, D.A., Clinton, J.D., Lapinski,
  J.S.}}} \bibinfo{year}{2012}\natexlab{}.
\newblock \bibinfo{title}{{A House Divided? Roll Calls, Polarization, and
  Policy Differences in the U.S. House, 1877-2011.}}
\newblock \bibinfo{howpublished}{{\textit{American Journal of Political
  Science}, 61 (3), pages 698-714}}.   (\bibinfo{year}{2012}).
\newblock


\bibitem[\protect\citeauthoryear{{Bianconi, G., Pin, P., Marsili,
  M.}}{{Bianconi, G., Pin, P., Marsili, M.}}{2009}]%
        {entropyMeasure}
\bibfield{author}{\bibinfo{person}{{Bianconi, G., Pin, P., Marsili, M.}}}
  \bibinfo{year}{2009}\natexlab{}.
\newblock \bibinfo{title}{{Assessing the relevance of node features for network
  structure}}.
\newblock \bibinfo{howpublished}{{\textit{Proceedings of the National Academy
  of Sciences}, 106, 11433}}.   (\bibinfo{year}{2009}).
\newblock


\bibitem[\protect\citeauthoryear{{Boxell, L., Gentzkow, M., Shapiro,
  J.M.}}{{Boxell, L., Gentzkow, M., Shapiro, J.M.}}{2017}]%
        {nberPolarization}
\bibfield{author}{\bibinfo{person}{{Boxell, L., Gentzkow, M., Shapiro, J.M.}}}
  \bibinfo{year}{2017}\natexlab{}.
\newblock \bibinfo{title}{Is the Internet Causing Political Polarization?
  Evidence from Demographics}.
\newblock \bibinfo{howpublished}{\textit{NBER Working Paper} No. 23258}.
  (\bibinfo{year}{2017}).
\newblock


\bibitem[\protect\citeauthoryear{{Conover, M.D., Ratkiewicz, J., Francisco, M.,
  Gonçalves, B., Flammini, A., Menczer, F.}}{{Conover, M.D., Ratkiewicz, J.,
  Francisco, M., Gonçalves, B., Flammini, A., Menczer, F.}}{2011}]%
        {polarizationTwitter}
\bibfield{author}{\bibinfo{person}{{Conover, M.D., Ratkiewicz, J., Francisco,
  M., Gonçalves, B., Flammini, A., Menczer, F.}}}
  \bibinfo{year}{2011}\natexlab{}.
\newblock \bibinfo{title}{{Political polarization on Twitter}}.
\newblock \bibinfo{howpublished}{{\textit{Proceedings of the 5th AAAI
  International Conference on Web and Social Media}}}.
  (\bibinfo{year}{2011}).
\newblock


\bibitem[\protect\citeauthoryear{{Cook, J., Lewandowsky, S.}}{{Cook, J.,
  Lewandowsky, S.}}{2011}]%
        {debunkingHandbook}
\bibfield{author}{\bibinfo{person}{{Cook, J., Lewandowsky, S.}}}
  \bibinfo{year}{2011}\natexlab{}.
\newblock \bibinfo{title}{{The Debunking Handbook}}.
\newblock \bibinfo{howpublished}{St. Lucia, Australia: University of
  Queensland}.   (\bibinfo{year}{2011}).
\newblock


\bibitem[\protect\citeauthoryear{{Eagle, N., Macy, M., Claxton, R.}}{{Eagle,
  N., Macy, M., Claxton, R.}}{2010}]%
        {networkDiversity}
\bibfield{author}{\bibinfo{person}{{Eagle, N., Macy, M., Claxton, R.}}}
  \bibinfo{year}{2010}\natexlab{}.
\newblock \bibinfo{title}{Network diversity and economic development}.
\newblock \bibinfo{howpublished}{\textit{Science}, 328, 1029-1031}.
  (\bibinfo{year}{2010}).
\newblock


\bibitem[\protect\citeauthoryear{{Fernbach, P. M., Rogers, T., Fox, C. R.,
  Sloman, S. A.}}{{Fernbach, P. M., Rogers, T., Fox, C. R., Sloman, S.
  A.}}{2013}]%
        {illusionUnderstanding}
\bibfield{author}{\bibinfo{person}{{Fernbach, P. M., Rogers, T., Fox, C. R.,
  Sloman, S. A.}}} \bibinfo{year}{2013}\natexlab{}.
\newblock \bibinfo{title}{{ Political Extremism is Supported by an Illusion of
  Understanding}}.
\newblock \bibinfo{howpublished}{{\textit{Psychological Science}, 24 (6), pages
  939-946}}.   (\bibinfo{year}{2013}).
\newblock


\bibitem[\protect\citeauthoryear{{Garimella, K., Morales, G.D.F., Gionis, A.,
  Mathioudakis, M.}}{{Garimella, K., Morales, G.D.F., Gionis, A., Mathioudakis,
  M.}}{2017}]%
        {reducingControversy}
\bibfield{author}{\bibinfo{person}{{Garimella, K., Morales, G.D.F., Gionis, A.,
  Mathioudakis, M.}}} \bibinfo{year}{2017}\natexlab{}.
\newblock \bibinfo{title}{{Reducing Controversy by Connecting Opposing Views}}.
\newblock \bibinfo{howpublished}{{\textit{Proceedings of the 10th ACM
  International Conference on Web Search and Data Mining}}}.
  (\bibinfo{year}{2017}).
\newblock


\bibitem[\protect\citeauthoryear{{Garimella, V., Weber, I.}}{{Garimella, V.,
  Weber, I.}}{2017}]%
        {twitterLongTerm}
\bibfield{author}{\bibinfo{person}{{Garimella, V., Weber, I.}}}
  \bibinfo{year}{2017}\natexlab{}.
\newblock \bibinfo{title}{{A Long-Term Analysis of Polarization on Twitter}}.
\newblock \bibinfo{howpublished}{{\textit{Proceedings of the 11th AAAI
  International Conference on Web and Social Media}}}.
  (\bibinfo{year}{2017}).
\newblock


\bibitem[\protect\citeauthoryear{{Gerber, A.S., Green, D.P.}}{{Gerber, A.S.,
  Green, D.P.}}{2012}]%
        {gerberGreen}
\bibfield{author}{\bibinfo{person}{{Gerber, A.S., Green, D.P.}}}
  \bibinfo{year}{2012}\natexlab{}.
\newblock \bibinfo{title}{{Field Experiments: Design, Analysis, and
  Interpretation}}.
\newblock \bibinfo{howpublished}{\textit{W.W. Norton \& Company}, York, PA.}.
  (\bibinfo{year}{2012}).
\newblock


\bibitem[\protect\citeauthoryear{{Haidt, J.}}{{Haidt, J.}}{2012}]%
        {righteousMind}
\bibfield{author}{\bibinfo{person}{{Haidt, J.}}}
  \bibinfo{year}{2012}\natexlab{}.
\newblock \bibinfo{title}{{The Righteous Mind: Why Good People Are Divided By
  Politics and Religion}}.
\newblock \bibinfo{howpublished}{\textit{Pantheon Books}, New York}.
  (\bibinfo{year}{2012}).
\newblock


\bibitem[\protect\citeauthoryear{{Iyengar, S., Sood, G., Lelkes, Y.}}{{Iyengar,
  S., Sood, G., Lelkes, Y.}}{2012}]%
        {affectNotIdeology}
\bibfield{author}{\bibinfo{person}{{Iyengar, S., Sood, G., Lelkes, Y.}}}
  \bibinfo{year}{2012}\natexlab{}.
\newblock \bibinfo{title}{{Affect, Not Ideology: A Social Identity Perspective
  on Polarization.}}
\newblock \bibinfo{howpublished}{{\textit{Public Opinion Quarterly}, 76 (3),
  pages 405-431}}.   (\bibinfo{year}{2012}).
\newblock


\bibitem[\protect\citeauthoryear{{Johnson, A.}}{{Johnson, A.}}{2014}]%
        {sharingEndorsement}
\bibfield{author}{\bibinfo{person}{{Johnson, A.}}}
  \bibinfo{year}{2014}\natexlab{}.
\newblock \bibinfo{title}{{The Ethics of Retweeting and Weather It Amounts To
  Endorsement.}}
\newblock \bibinfo{howpublished}{{NPR}}.   (\bibinfo{year}{2014}).
\newblock


\bibitem[\protect\citeauthoryear{{Lelkes, Y., Sood, G., Iyengar, S.}}{{Lelkes,
  Y., Sood, G., Iyengar, S.}}{1971}]%
        {broadbandPolarization}
\bibfield{author}{\bibinfo{person}{{Lelkes, Y., Sood, G., Iyengar, S.}}}
  \bibinfo{year}{1971}\natexlab{}.
\newblock \bibinfo{title}{The hostile audience: The effect of access to
  broadband internet on partisan affect}.
\newblock \bibinfo{howpublished}{\textit{American Journal of Political
  Science}, 61 (1): 5-20}.   (\bibinfo{year}{1971}).
\newblock


\bibitem[\protect\citeauthoryear{{Matias, N., Szalavitz, S., Zuckerman,
  E.}}{{Matias, N., Szalavitz, S., Zuckerman, E.}}{2017}]%
        {followBias}
\bibfield{author}{\bibinfo{person}{{Matias, N., Szalavitz, S., Zuckerman, E.}}}
  \bibinfo{year}{2017}\natexlab{}.
\newblock \bibinfo{title}{{FollowBias: Supporting Behavior Change toward Gender
  Equality by Networked Gatekeepers on Social Media}}.
\newblock \bibinfo{howpublished}{{\textit{Proceedings of the 2017 ACM
  Conference on Computer Supported Cooperative Work and Social Computing}}}.
  (\bibinfo{year}{2017}).
\newblock


\bibitem[\protect\citeauthoryear{{Mislove, A., Lehmann, S.; Ahn, Y.; Onnela,
  J.; and Rosenquist, J.}}{{Mislove, A., Lehmann, S.; Ahn, Y.; Onnela, J.; and
  Rosenquist, J.}}{2011}]%
        {twitterNonRepresentative}
\bibfield{author}{\bibinfo{person}{{Mislove, A., Lehmann, S.; Ahn, Y.; Onnela,
  J.; and Rosenquist, J.}}} \bibinfo{year}{2011}\natexlab{}.
\newblock \bibinfo{title}{Understanding the Demographics of Twitter Users}.
\newblock \bibinfo{howpublished}{\textit{Proceedings of the 5th AAAI
  International Conference on Web and Social Media}}.   (\bibinfo{year}{2011}).
\newblock


\bibitem[\protect\citeauthoryear{{Munger, K.}}{{Munger, K.}}{2016}]%
        {mungerOnlineRacism}
\bibfield{author}{\bibinfo{person}{{Munger, K.}}}
  \bibinfo{year}{2016}\natexlab{}.
\newblock \bibinfo{title}{{Tweetment Effects on the Tweeted: Experimentally
  Reducing Racist Harassment}}.
\newblock \bibinfo{howpublished}{{\textit{Political Behavior}, 61 (3), pages
  698-714}}.   (\bibinfo{year}{2016}).
\newblock


\bibitem[\protect\citeauthoryear{{Munson, S.A., Lee, S.Y., Resnick,
  P.}}{{Munson, S.A., Lee, S.Y., Resnick, P.}}{2013}]%
        {munsonExtension}
\bibfield{author}{\bibinfo{person}{{Munson, S.A., Lee, S.Y., Resnick, P.}}}
  \bibinfo{year}{2013}\natexlab{}.
\newblock \bibinfo{title}{Encouraging Reading of Diverse Political Viewpoints
  with a Browser Widget}.
\newblock \bibinfo{howpublished}{\textit{Proceedings of the 7th AAAI
  International Conference on Web and Social Media}}.   (\bibinfo{year}{2013}).
\newblock


\bibitem[\protect\citeauthoryear{{Nikolov, D., Fregolente, D., Flammini, D.,
  Menczer, F.}}{{Nikolov, D., Fregolente, D., Flammini, D., Menczer,
  F.}}{2015}]%
        {measuringBubbles}
\bibfield{author}{\bibinfo{person}{{Nikolov, D., Fregolente, D., Flammini, D.,
  Menczer, F.}}} \bibinfo{year}{2015}\natexlab{}.
\newblock \bibinfo{title}{{Measuring Online Social Bubbles}}.
\newblock \bibinfo{howpublished}{{\textit{PeerJ Computer Science}, 1:e38}}.
  (\bibinfo{year}{2015}).
\newblock


\bibitem[\protect\citeauthoryear{{Pew Research Center}}{{Pew Research
  Center}}{2014}]%
        {pewPolarization}
\bibfield{author}{\bibinfo{person}{{Pew Research Center}}.}
  \bibinfo{year}{2014}\natexlab{}.
\newblock \bibinfo{title}{{Political Polarization in the American Public}}.
\newblock   (\bibinfo{year}{2014}).
\newblock


\bibitem[\protect\citeauthoryear{{Sunstein, C.R.}}{{Sunstein, C.R.}}{2002}]%
        {sunsteinPolarization}
\bibfield{author}{\bibinfo{person}{{Sunstein, C.R.}}}
  \bibinfo{year}{2002}\natexlab{}.
\newblock \bibinfo{title}{The Law of Group Polarization}.
\newblock \bibinfo{howpublished}{\textit{Journal of Political Philosophy}, 10,
  175-195}.   (\bibinfo{year}{2002}).
\newblock


\bibitem[\protect\citeauthoryear{{Vijayaraghavan, P., Vosoughi, S., Roy,
  D.}}{{Vijayaraghavan, P., Vosoughi, S., Roy, D.}}{2016}]%
        {lsmElectome}
\bibfield{author}{\bibinfo{person}{{Vijayaraghavan, P., Vosoughi, S., Roy,
  D.}}} \bibinfo{year}{2016}\natexlab{}.
\newblock \bibinfo{title}{Automatic Detection and Categorization of
  Election-Related Tweets}.
\newblock \bibinfo{howpublished}{\textit{Proceedings of the 10th AAAI
  International Conference on Web and Social Media}}.   (\bibinfo{year}{2016}).
\newblock


\bibitem[\protect\citeauthoryear{{Vijayaraghavan, P., Vosoughi, S., Roy,
  D.}}{{Vijayaraghavan, P., Vosoughi, S., Roy, D.}}{2017}]%
        {twitterDemos}
\bibfield{author}{\bibinfo{person}{{Vijayaraghavan, P., Vosoughi, S., Roy,
  D.}}} \bibinfo{year}{2017}\natexlab{}.
\newblock \bibinfo{title}{{Twitter Demographic Classification Using Deep
  Multi-modal Multi-task Learning}}.
\newblock \bibinfo{howpublished}{{\textit{Proceedings of the 55th Annual
  Meeting of the Association for Computational Linguistics}, pages 478-483}}.
  (\bibinfo{year}{2017}).
\newblock


\bibitem[\protect\citeauthoryear{{Westwood, S.J., Iyengar, S., Walgrave, S.,
  Leonisio, R., Miller, L., Strijbis, O.}}{{Westwood, S.J., Iyengar, S.,
  Walgrave, S., Leonisio, R., Miller, L., Strijbis, O.}}{2017}]%
        {primacyPartyism}
\bibfield{author}{\bibinfo{person}{{Westwood, S.J., Iyengar, S., Walgrave, S.,
  Leonisio, R., Miller, L., Strijbis, O.}}} \bibinfo{year}{2017}\natexlab{}.
\newblock \bibinfo{title}{{The tie that divides: Cross-national evidence of the
  primacy of partyism.}}
\newblock \bibinfo{howpublished}{{\textit{European Journal of Political
  Research}}}.   (\bibinfo{year}{2017}).
\newblock


\bibitem[\protect\citeauthoryear{{Yardi, S., boyd, d.}}{{Yardi, S., boyd,
  d.}}{2010}]%
        {yardiboyd}
\bibfield{author}{\bibinfo{person}{{Yardi, S., boyd, d.}}}
  \bibinfo{year}{2010}\natexlab{}.
\newblock \bibinfo{title}{Dynamic debates: An analysis of group polarization
  over time on twitter}.
\newblock \bibinfo{howpublished}{\textit{Bulletin of Science, Technology and
  Society}, 20:1-8}.   (\bibinfo{year}{2010}).
\newblock


\end{thebibliography}

\appendix
\section{Appendix}
To recruit study participants (Section~\ref{sec:recruitment}) we sent the following message as a Twitter Direct Message: \\

\begin{quote}

\textit{We are a team of researchers at MIT analyzing political polarization on social media platforms. We have created a tool called "Social Mirror" that enables you to interactively explore the politically-active parts of your social network on Twitter. We are inviting a small group of people to try out the tool, which you can find here:}\\

\url{https://socialmirror.media.mit.edu}\\

\textit{We've found that many politically-active Twitter users embed themselves in echo chambers, surrounded by people whose political perspectives reflect their own. As a well-functioning democracy requires its citizens to actively engage with different viewpoints, we are interested in learning more about how we might help mitigate online echo chambers.}\\

\textit{If you have any ideas, questions, or feedback, please message us at socialmirror@media.mit.edu. We look forward to hearing from you!}
\end{quote}


\end{document}